\documentclass[pre, twocolumn, showpacs, showkeys, amsmath, amssymb]{revtex4}

\usepackage{graphicx}

\begin{document}

\title{Evolutionary dynamics of adult stem cells:  Comparison of 
random and immortal strand co-segregation mechanisms}

\author{Emmanuel Tannenbaum}
\email{etannenb@fas.harvard.edu}
\affiliation{Harvard University, Cambridge, MA 02138}
\author{James L. Sherley}
\email{jsherley@mit.edu}
\affiliation{Biological Engineering Division, Massachussetts Institute of
Technology, Cambridge, MA 02139}
\author{Eugene I. Shakhnovich}
\email{eugene@belok.harvard.edu}
\affiliation{Department of Chemistry and Chemical Biology, Harvard University,
Cambridge, MA 02138}

\begin{abstract}

This paper develops a point-mutation model describing the evolutionary
dynamics of a population of adult stem cells.  Such a model may prove 
useful for quantitative studies of tissue aging and the emergence of
cancer.  We consider two modes of chromosome segregation:  (1) Random 
segregation, where the daughter chromosomes of a given parent chromosome 
segregate randomly into the stem cell and its differentiating sister cell.  
(2) ``Immortal DNA strand'' co-segregation, for which the stem cell retains 
the daughter chromosomes with the oldest parent strands.  Immortal strand 
co-segregation is a mechanism, originally proposed by Cairns (J. Cairns, 
{\it Nature} {\bf 255}, 197 (1975)), by which stem cells preserve the 
integrity of their genomes.  For random segregation, we develop an ordered 
strand pair formulation of the dynamics, analogous to the ordered strand pair 
formalism developed for quasispecies dynamics involving semiconservative 
replication with imperfect lesion repair (in this context, lesion repair is
taken to mean repair of postreplication base-pair mismatches).  Interestingly, 
a similar formulation is possible with immortal strand co-segregation, despite 
the fact that this segregation mechanism is age-dependent.  From our model we 
are able to mathematically show that, when lesion repair is imperfect, then 
immortal strand co-segregation leads to better preservation of the stem cell 
lineage than random chromosome segregation.  Furthermore, our model allows us 
to estimate the optimal lesion repair efficiency for preserving an adult stem 
cell population for a given period of time.  For human stem cells, we obtain 
that mispaired bases still present after replication and cell division should 
be left untouched, to avoid potentially fixing a mutation in both DNA strands. 

\end{abstract}



\maketitle

\section{Introduction}

The generation and maintenance of tissues in mammals is currently a topic of 
intense investigation by experimental and theoretical biologists.  Besides its 
intrinsic scientific interest, an understanding of tissue cell kinetics,
architecture, and development has important implications for
aging and cancer.  

In vertebrate animals, many tissues and organs are generated by what
are known as adult (or equivalently, somatic) stem cells.  Adult stem
cells are rare, undifferentiated cells that divide asymmetrically to 
renew differentiated cells in adult tissues.  They divide to produce the
original stem cell, and a differentiating progeny cell.  The differentiating
progeny cell then proceeds through a series of division and differentiation 
steps (see Figure 1), to produce a large collection of mature tissue cells.
  
\begin{figure}
\includegraphics[width = 0.9\linewidth]{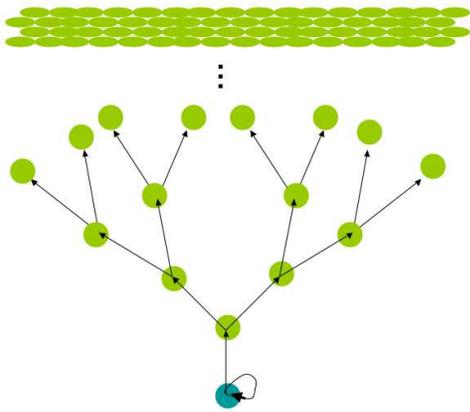}
\caption{(Color online) Generation of differentiated tissue cells (green) from 
an adult stem cell (blue).}
\end{figure}
                                                
At this point, it is not clear how adult stem cells emerge in 
multicellular organisms, nor is it known how this method of generating tissue 
cells evolved.  Nevertheless, it is believed that this mechanism
may serve to delay the emergence of cancer in mammals.  

Mature skin cells, for example, are continually regenerated by adult stem 
cells.  The tissue cells, after undergoing a prespecified number of divisions, 
cease dividing (a process known as terminal differentiation), and are 
eventually shed.  Thus, any potentially cancerous mutation in differentiated
skin tissue cells will eventually leave the body, thereby reducing the risk of 
skin cancer.  

In order to effectively reduce mutation rates, however, there must exist
a mechanism or collection of mechanisms that protect the genetic integrity
of the adult stem cell population.  Otherwise, because adult stem cells
are long-lived in the body, they will eventually accumulate a sufficient number
of mutations to become cancerous, or become genetically inferior stem cells.

One important mechanism by which adult stem cells protect the integrity of
their genomes is through a form of asymmetric chromosome segregation
during cell division, known as {\it immortal DNA strand} co-segregation.
The immortal strand hypothesis was originally proposed by Cairns 
\cite{CAIRNS1}.  It states that when an adult stem cell divides to form
a stem cell and a differentiating tissue cell, the stem cell retains
the chromosomes with the oldest DNA strands of the genome (see Figure 2).  
Presumably, the oldest DNA strands of the genome provide the most accurate 
template for daughter strand synthesis, and hence their preferential 
segregation into the adult stem cells ensures optimal maintenance of stem cell 
genetic integrity and overall tissue health.

\begin{figure}
\includegraphics[width = 0.9\linewidth]{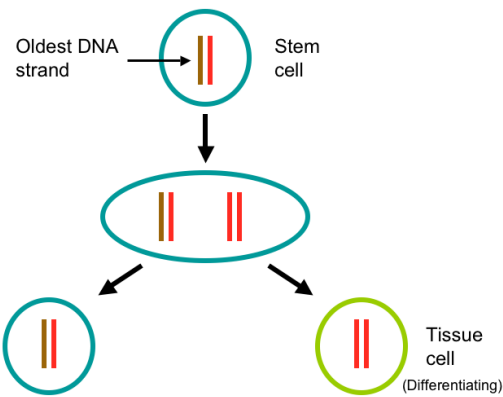}
\caption{(Color online) Illustration of immortal DNA strand chromosome 
segregation.}
\end{figure}

The immortal strand mechanism was recently confirmed experimentally 
\cite{SHERLEY, POTTEN1}.  The confirmation of this segregation mechanism
has motivated the authors to develop a mathematical model describing
the evolutionary dynamics of a population of adult stem cells.

We are interested in three aspects of stem cell evolutionary 
dynamics:  First of all, we seek to develop a set of ordinary differential 
equations describing the evolutionary dynamics of a population of adult stem 
cells.  This is done in the following section.  For simplicity, we assume 
an infinite population, continuous time model.  While strictly speaking this 
is not correct, stochastic simulations show good agreement already at 
populations with as few as $ 10,000 $ stem cells.

Second, we wish to rigorously show that immortal strand co-segregation is 
necessary to preserve the stem cell lineage.  Immortal strand co-segregation 
can only provide an advantage, however, if, during a process known
as {\it lesion repair}, not all postreplication DNA mismatches are corrected
Otherwise, daughter-strand synthesis errors can become fixed as mutations in 
both parent and daughter strands, thereby eliminating the advantage of keeping 
the oldest template strand in the stem cell \cite{VOET, SEMICONSERV, BRUMER, 
LESFULL}.

Finally, because a high lesion repair efficiency reduces the overall mutation
rate, while low lesion repair efficiency preserves the information in
the parent strand, there is an optimal lesion repair efficiency for 
maximally preserving the stem cell lineage for a given period of time.  In our
case, the period of time of interest is a human lifetime, which we take
to be on the order of $ 80 $ years.

In the following section, we derive the finite sequence length equations
describing the evolutionary dynamics of adult stem cells, for the cases
of random segregation versus immortal strand co-segregation.  In particular, 
we develop an ordered strand pair formulation of the dynamics, analogous to the
ordered strand pair formulation of the quasispecies equations for
semiconservative replication with imperfect lesion repair \cite{SEMICONSERV,
BRUMER, LESFULL}.  For random segregation, the equations derived are similar 
to the corresponding quasispecies equations.  For immortal strand 
co-segregation, the equations are qualitatively different.  Nevertheless, 
despite the age-dependence of the chromosome segregation mechanism, for 
immortal strand co-segregation it is still possible to develop an ordered 
strand pair formulation of the dynamics.

In Section III, we derive the infinite sequence length form of the evolutionary
dynamics equations, for a class of fitness landscapes defined by a
master genome.  These equations are analogous to the equations developed
for semiconservative replication with imperfect lesion repair \cite{LESFULL}.
We then proceed to obtain the system of differential equations governing
the decay of the stem cell population with the master-genome genotype.

We continue in Section IV, where we use the master-genome equations to
determine the optimal lesion repair efficiency for preserving the stem
cell lineage for a given amount of time.  In particular, we show that
lesion repair should be turned off in stem cells.  That is, postreplication
DNA mismatches should be left uncorrected in stem cells.  

We conclude in Section V with a summary of our results, and plans for future
research.

\section{Derivation of the Finite Sequence Length Equations}

\subsection{Definitions}

We consider a population of $ N_S $ replicating adult stem cells.  As is 
illustrated in Figure 1, each of these stem cells generates a lineage of 
differentiated tissue cells.  

We assume that each stem cell has a genome consisting of a single, 
double-stranded DNA molecule.  A given genome may then be given by the
set $ \{\sigma, \sigma'\} $, where $ \sigma $ and $ \sigma' $ denote
the two strands.  In principle, DNA consists of two antiparallel,
complementary strands.  Thus, a genome of length $ L $ should consist of 
the strands $ \sigma $ and its complement $ \bar{\sigma} $, where $ \sigma =
b_1 \dots b_L \Leftrightarrow \bar{\sigma} = \bar{b}_L \dots \bar{b}_1 $
($ \bar{b}_i $ denotes the complement of $ b_i $.  For the four bases
used in DNA, complementary is defined by the Watson-Crick pairs
Adenine:Thymine (A:T) and Guanine:Cytosine (G:C).  See 
Figure 1 in \cite{SEMICONSERV}).  However, due to mutations, it is
possible that the two strands of a given genome are not perfectly 
complementary, and so we have to relax this restriction.
 
We also assume first-order growth, so that with each genome $ \{\sigma,
\sigma'\} $ is associated a first-order growth rate constant
$ \kappa_{\{\sigma, \sigma'\}} $.  The collection of all first-order
growth rate constants is known as the {\it fitness landscape}.  For 
simplicity, we assume in this paper a {\it static}, or time-independent
landscape.

As with all cells with double-stranded DNA genomes, we assume semiconservative
replication, where the genome of each cell unzips to form two strands, each
of which serves as a template for the synthesis of the complementary 
daughter strands.  The end result is two new daughter genomes, one of
which is retained by the stem cell, while the other becomes the genome 
of the differentiating sister.  When genome $ \{\sigma, \sigma'\} $ 
replicates, then we assume that with daughter strand synthesis is associated 
a per-base mismatch probability of $ \epsilon_{\{\sigma, \sigma'\}} $.  

After replication is complete, and stem cell division has occurred, there may 
still be some errors in the daughter strands which were missed by various 
error-correction mechanisms (DNA polymerase proofreading and mismatch repair). 
These mismatches result in lesions along the DNA chain, which may be recognized
and repaired by various maintenance enzymes in the cell.  It should be noted
that in this case, the cell cannot distinguish between parent and
daughter strands (which it does during daughter strand synthesis).  Thus,
a given error in the daughter strand has a $ 50\% $ probability of being
corrected, but it also has a $ 50\% $ probability of being communicated
to the parent strand.  When this happens, the mutation is said to be
{\it fixed} in the genome.  Lesion repair is generally not perfect, and so we 
assume that when genome $ \{\sigma, \sigma'\} $ replicates, a 
postreplication mismatch in the resulting daughter genomes is repaired with 
probability $ \lambda_{\{\sigma, \sigma'\}} $.

Errors during daughter strand synthesis and lesion repair result in 
a probability distribution for the possible daughter genome which can
be generated from a given parent strand.  Thus, we define 
$ p((\sigma'', \sigma'''), \{\sigma, \sigma'\}) $ to be the probability
that parent strand $ \sigma'' $, as part of genome $ \{\sigma'', \sigma'''\} $,
forms the daughter genome $ \{\sigma, \sigma'\} $.  

We may also note that $ \sigma'' $ can form $ \{\sigma, \sigma'\} $ by
either becoming $ \sigma $, with daughter strand $ \sigma' $, or
$ \sigma' $, with daughter strand $ \sigma $.  The probability of
the former process is denoted by $ p((\sigma'', \sigma'''), 
(\sigma, \sigma')) $, and the probability of the latter process is
denoted by $ p((\sigma'', \sigma'''), (\sigma', \sigma)) $.  Note
that if $ \sigma \neq \sigma' $, then 
$ p((\sigma'', \sigma'''), \{\sigma, \sigma'\}) =
p((\sigma'', \sigma'''), (\sigma, \sigma')) + 
p((\sigma'', \sigma'''), (\sigma', \sigma)) $, while
$ p((\sigma'', \sigma'''), \{\sigma, \sigma\}) = 
p((\sigma'', \sigma'''), (\sigma, \sigma)) $. 
An expression for $ p((\sigma'', \sigma'''), 
(\sigma, \sigma')) $ was derived in \cite{LESFULL}.  

Finally, because stem cell division (more properly, asymmetric
self-renewal) results in a constant value for $ N_S $, it is equivalent to 
look at population fractions.  We therefore define $ x_{\{\sigma, \sigma'\}} $ 
to be the fraction of the stem cell population (at a given time $ t $) with 
genome $ \{\sigma, \sigma'\} $.  

For immortal strand co-segregation, the preceding definitions need to be
somewhat modified, since we need to also keep track of the ages of the strands.
To this end, we let $ \sigma^{(T)} $ denote a strand which has been the
template (parent) strand at least once, while $ \sigma^{(N)} $ denotes
a strand which has never been the template for the synthesis of a daughter
strand.  For immortal strand co-segregation, then, we consider genomes
of the form $ \{\sigma^{(N)}, \sigma'^{(N)}\} $ and $ \{\sigma^{(T)},
\sigma'^{(N)}\} $.  We do not consider genomes of the form
$ \{\sigma^{(T)}, \sigma'^{(T)}\} $, since, if our population initially
consists of genomes which have never been involved in daughter strand 
synthesis, then such genomes can never appear in the population.  The reason
for this is that when a parent strand serves as the template for daughter
strand synthesis, then it should be clear that the daughter strand
automatically receives the ``N'' designation.  Thus, two ``T'' strands
can never be paired with one another.

\subsection{Random segregation}

For random chromosome segregation, each of the parent strands of a replicating
genome has an equal probability of becoming incorporated into the
stem cell.  The random segregation equations are then given by,
\begin{eqnarray}
\frac{d x_{\{\sigma, \sigma'\}}}{dt} 
& = &
-\kappa_{\{\sigma, \sigma'\}} x_{\{\sigma, \sigma'\}} \nonumber \\
&   &
+ \frac{1}{2}\sum_{\{\sigma'', \sigma'''\}} 
\kappa_{\{\sigma'', \sigma'''\}} x_{\{\sigma'', \sigma'''\}}
\times \nonumber \\
&   &
[p((\sigma'', \sigma'''), \{\sigma, \sigma'\}) + 
 p((\sigma''', \sigma''), \{\sigma, \sigma'\})]
\nonumber \\
\end{eqnarray}
The term $ -\kappa_{\{\sigma, \sigma'\}} x_{\{\sigma, \sigma'\}} $
arises from the observation that, in semiconservative replication, 
the separation of the parent strands corresponds to the effective destruction
of the original genome.  The second term gives the rate at which 
$ \{\sigma, \sigma'\} $ is produced, due to replication and mutation, by
all genomes in the population.  The factor of $ 1/2 $ arises because
for random chromosome segregation, both parent strands $ \sigma'' $ and 
$ \sigma''' $ of a replicating genome $ \{\sigma'', \sigma'''\} $ have an
equal probability of being retained by the stem cell.

The above equations are fairly cumbersome for direct analysis, since the
dynamics occurs over a space of double-stranded genomes.  If the strands
are completely correlated, so that in a genome $ \{\sigma, \sigma'\} $ 
we always have $ \sigma' = \bar{\sigma} $, then following the derivation 
in \cite{SEMICONSERV}, it is possible to convert the dynamics over the 
space of double-stranded genomes into an equivalent dynamics
over the space of single strands.  This conversion is not possible when
the assumption of complementarity does not hold.  Nevertheless, following
the derivation in \cite{LESFULL}, we can convert the dynamics over
the space of double-stranded genomes into an equivalent dynamics
over the space of ordered strand pairs.  Specifically, given
some genome $ \{\sigma, \sigma'\} $, define,
\begin{eqnarray}
y_{(\sigma, \sigma')} = y_{(\sigma', \sigma)} =
\left\{ \begin{array}{cc}
        \frac{1}{2} x_{\{\sigma, \sigma'\}} &
        \mbox{if $ \sigma \neq \sigma' $} \\
        x_{\{\sigma, \sigma'\}} &
        \mbox{if $ \sigma = \sigma' $}
        \end{array}
\right.       
\end{eqnarray}
Furthermore, define an ordered strand pair
fitness landscape via $ \kappa_{(\sigma, \sigma')} = \kappa_{(\sigma',
\sigma)} = \kappa_{\{\sigma, \sigma'\}} $.  The random segregation
equations then become,
\begin{eqnarray}
\frac{d y_{(\sigma, \sigma')}}{dt}
& = &
-\kappa_{(\sigma, \sigma')} y_{(\sigma, \sigma')} \nonumber \\
&   &
+ \frac{1}{2} \sum_{(\sigma'', \sigma''')}
\kappa_{(\sigma'', \sigma''')} y_{(\sigma'', \sigma''')}
\nonumber \\
&   &
[p((\sigma'', \sigma'''), (\sigma, \sigma')) + 
 p((\sigma'', \sigma'''), (\sigma', \sigma))]
\nonumber \\
\end{eqnarray}

\subsection{Immortal strand co-segregation}

To derive the evolutionary dynamics for a stem cell population replicating
with immortal strand co-segregation, we have to take into account the ages
of the strands.  In this case, we have to separately derive the dynamics for
genomes where neither strand has been used as a template for daughter
strand synthesis, and where one of the strands has been used as a
template for daughter strand synthesis.  The resulting system of equations
is given by, 
\begin{eqnarray}
\frac{d x_{\{\sigma^{(N)}, \sigma'^{(N)}\}}}{dt}
& = &
-\kappa_{\{\sigma, \sigma'\}} x_{\{\sigma^{(N)}, \sigma'^{(N)}\}} \nonumber \\
\frac{d x_{\{\sigma^{(T)}, \sigma'^{(N)}\}}}{dt}
& = &
-\kappa_{\{\sigma, \sigma'\}} x_{\{\sigma^{(T)}, \sigma'^{(N)}\}} \nonumber \\
&   &
+ \frac{1}{2} \sum_{\{\sigma''^{(N)}, \sigma'''^{(N)}\}}
\kappa_{\{\sigma'', \sigma'''\}} x_{\{\sigma''^{(N)}, \sigma'''^{(N)}\}}
\times \nonumber \\
&   &
[p((\sigma'', \sigma'''), (\sigma, \sigma')) + 
 p((\sigma''', \sigma''), (\sigma, \sigma'))] \nonumber \\
&   &
+ \sum_{\{\sigma''^{(T)}, \sigma'''^{(N)}\}}
\kappa_{\{\sigma'', \sigma'''\}} x_{\{\sigma''^{(T)}, \sigma'''^{(N)}\}}
\times \nonumber \\
&   &
p((\sigma'', \sigma'''), (\sigma, \sigma'))
\end{eqnarray}
Note that genomes of the form $ \{\sigma^{(N)}, \sigma'^{(N)}\} $ cannot
be produced via replication, since replication occurs via a parent strand which
has then been used as a template for daughter strand synthesis at least
once.  

Note also that when a genome $ \{\sigma''^{(N)}, \sigma'''^{(N)}\} $
replicates, strands $ \sigma'' $ and $ \sigma''' $ have an equal 
probability of being retained by the stem cell.  Of course, when
a genome $ \{\sigma''^{(T)}, \sigma'''^{(N)}\} $ replicates, then
it is strand $ \sigma'' $ that is retained by the stem cell.

Finally, note in the second equation that we are not considering
probabilities $ p((\sigma'', \sigma'''), \{\sigma, \sigma'\}) $, but
rather probabilities $ p((\sigma'', \sigma'''), (\sigma, \sigma')) $.
The reason for this is that in considering the production of genome
$ \{\sigma^{(T)}, \sigma'^{(N)}\} $, strand $ \sigma $ is explicitly
marked as the template strand, while strand $ \sigma' $ is explicitly
marked as the newly synthesized daughter strand.  Therefore, to form
$ \{\sigma^{(T)}, \sigma'^{(N)}\} $, it is clear that the parent 
(template) strand $ \sigma'' $ must become $ \sigma $, with daughter
strand $ \sigma' $.

As with the random segregation equations, we may define a equivalent
dynamics over the space of ordered strand pairs.  We do this in two
steps.  First, define,
\begin{eqnarray}
y_{(\sigma^{(N)}, \sigma'^{(N)})} = y_{(\sigma'^{(N)}, \sigma^{(N)})} =
\left\{ \begin{array}{cc}
        \frac{1}{2} x_{\{\sigma^{(N)}, \sigma'^{(N)}\}} &
        \mbox{if $ \sigma \neq \sigma' $} \\
        x_{\{\sigma^{(N)}, \sigma'^{(N)}\}} &
        \mbox{if $ \sigma = \sigma' $}
        \end{array}
\right.       
\end{eqnarray}
and
\begin{equation}
y_{(\sigma^{(T)}, \sigma'^{(N)})} = x_{\{\sigma^{(T)}, \sigma'^{(N)}\}} 
\end{equation}
The ordered strand pair fitness landscape is defined as for random
segregation.  The result is the transformed system of equations,
\begin{eqnarray}
\frac{d y_{(\sigma^{(N)}, \sigma'^{(N)})}}{dt}
& = &
-\kappa_{(\sigma, \sigma')} y_{(\sigma^{(N)}, \sigma'^{(N)})} \nonumber \\
\frac{d y_{(\sigma^{(T)}, \sigma'^{(N)})}}{dt}
& = &
-\kappa_{(\sigma, \sigma')} y_{(\sigma^{(T)}, \sigma'^{(N)})} \nonumber \\
&   &
+ \sum_{(\sigma''^{(N)}, \sigma'''^{(N)})}
\kappa_{(\sigma'', \sigma''')} y_{(\sigma''^{(N)}, \sigma'''^{(N)})}
\times \nonumber \\
&   &
p((\sigma'', \sigma'''), (\sigma, \sigma')) \nonumber \\
&   &
+ \sum_{(\sigma''^{(T)}, \sigma'''^{(N)})}
\kappa_{(\sigma'', \sigma''')} y_{(\sigma''^{(T)}, \sigma'''^{(N)})}
\times \nonumber \\
&   &
p((\sigma'', \sigma'''), (\sigma, \sigma'))
\end{eqnarray}
The key equality to note in deriving the transformed dynamics is,
\begin{eqnarray}
&   &
\sum_{\{\sigma''^{(N)}, \sigma'''^{(N)}\}}
\kappa_{\{\sigma'', \sigma'''\}} x_{\{\sigma''^{(N)}, \sigma'''^{(N)}\}}
\times \nonumber \\
&   &
[p((\sigma'', \sigma'''), (\sigma, \sigma')) + 
 p((\sigma''', \sigma''), (\sigma, \sigma'))]
\nonumber \\
&   &
= 2 \sum_{\{\sigma''^(N), \sigma'''^{(N)}\}, \sigma'' \neq \sigma'''}
\times \nonumber \\
&   &
[\kappa_{(\sigma'', \sigma''')} y_{(\sigma''^{(N)}, \sigma'''^{(N)})}
p((\sigma'', \sigma'''), (\sigma, \sigma')) 
\nonumber \\
&   &
+ \kappa_{(\sigma''', \sigma'')} y_{(\sigma'''^{(N)}, \sigma''^{(N)})}
p((\sigma''', \sigma''), (\sigma, \sigma'))]
\nonumber \\
&   &
+ 2 \sum_{\{\sigma''^{(N)}, \sigma''^{(N)}\}}
\kappa_{(\sigma'', \sigma'')} y_{(\sigma''^{(N)}, \sigma''^{(N)})}
p((\sigma'', \sigma''), (\sigma, \sigma'))
\nonumber \\
&   &
= 2 \sum_{(\sigma''^{(N)}, \sigma'''^{(N)})} 
\kappa_{(\sigma'', \sigma''')} y_{(\sigma''^{(N)}, \sigma'''^{(N)})}
p((\sigma'', \sigma'''), (\sigma, \sigma')) \nonumber \\
\end{eqnarray}

Finally, if we define $ y_{(\sigma, \sigma')} = 
y_{(\sigma^{(N)}, \sigma'^{(N)})} + y_{(\sigma^{(T)}, \sigma^{(N)})} $,
then we obtain,
\begin{eqnarray}
\frac{d y_{(\sigma, \sigma')}}{dt}
& = &
-\kappa_{(\sigma, \sigma')} y_{(\sigma, \sigma')} 
\nonumber \\
&   &
+ \sum_{(\sigma'', \sigma''')}
\kappa_{(\sigma'', \sigma''')} y_{(\sigma'', \sigma''')}
p((\sigma'', \sigma'''), (\sigma, \sigma')) \nonumber \\
\end{eqnarray}

Note that the ordered strand pair population fractions are defined 
somewhat differently for immortal and random chromosome segregation.
For random chromosome segregation, the age of the strands is irrelevant
to the division kinetics.  Given a genome $ \{\sigma, \sigma'\} $,
there is no canonical ordering of the strands $ \sigma $ and $ \sigma' $.
If $ \sigma \neq \sigma' $, then the ordered pairs $ (\sigma, \sigma')
$ and $ (\sigma', \sigma) $ should receive identical contributions from
the genome $ \{\sigma, \sigma'\} $.

For immortal strand co-segregation, the above argument holds for genomes
of the form $ \{\sigma^{(N)}, \sigma'^{(N)}\} $.  However, for genomes
of the form $ \{\sigma^{(T)}, \sigma'^{(N)}\} $, a canonical ordering
of the strands exists.  Namely, we place the older strand before the
younger in the ordered strand pair representation.  This means that, for 
immortal strand co-segregation, we may regard $ y_{(\sigma, \sigma')} $ 
to be the total fraction of stem cells with template strand $ \sigma $ 
and daughter strand $ \sigma' $.  The only potential problem with this 
interpretation is the inclusion of $ y_{(\sigma^{(N)}, \sigma'^{(N)})} $ 
as part of this population fraction.  However, this may be resolved by 
noting that while $ \{\sigma^{(N)}, \sigma'^{(N)}\} $ has not yet undergone 
a replication cycle, when it does, either $ \sigma^{(N)} $ or 
$ \sigma'^{(N)} $ will be segregated into the original stem cell.  
Therefore, we may effectively {\it preassign} a ``T'' designation to 
either $ \sigma  $ or $ \sigma' $.  If $ \sigma = \sigma' $, then 
$ \sigma $ is the preassigned template strand for all genomes, while if 
$ \sigma \neq \sigma' $, then $ \sigma $ is the preassigned template strand 
for half of the genomes.  This interpretation for $ y_{(\sigma, \sigma')} $ 
is consistent with the definition for $ y_{(\sigma, \sigma')} $ 
($ 1/2 x_{\{\sigma^{(N)}, \sigma'^{(N)}\}} + x_{\{\sigma^{(T)}, 
\sigma'^{(N)}\}} $ for $ \sigma \neq \sigma' $, and $ x_{\{\sigma^{(N)}, 
\sigma^{(N)}\}} + x_{\{\sigma^{(T)}, \sigma^{(N)}\}} $ if 
$ \sigma = \sigma' $).  

In contrast to random chromosome segregation, for immortal strand 
co-segregation it is not generally true that $ y_{(\sigma', \sigma)} = 
y_{(\sigma, \sigma')} $.  The reason for this is that in the case of 
$ (\sigma, \sigma') $, $ \sigma $ is the template strand which has been 
present through all stem cell divisions (though perhaps mutated to something 
different from the original strand).  In the case of $ (\sigma', \sigma) $, it 
is $ \sigma' $ that has remained in the stem cell.  If $ \sigma $ and 
$ \sigma' $ are different, there is no reason to expect an identical 
evolutionary pathway for the two strands, hence it is incorrect to assume 
that $ y_{(\sigma, \sigma')} = y_{(\sigma', \sigma)} $. 

\subsection{Equivalence of random and immortal strand co-segregation when
lesion repair is perfectly efficient}

Under very general conditions, it is possible to show that when lesion
repair is perfect, then random and immortal strand co-segregation yield
identical stem cell dynamics.  We need only make the following assumptions:
(1)  For any ordered strand pair $ (\sigma, \sigma') $, we have 
$ \kappa_{(\bar{\sigma}, \bar{\sigma}')} = \kappa_{(\sigma, \sigma')} $.
(2)  For any two ordered strand pairs $ (\sigma, \sigma') $ and
$ (\sigma'', \sigma''') $, we have $ p((\sigma'', \sigma'''), 
(\sigma, \sigma')) = p((\bar{\sigma}'', \bar{\sigma}'''), 
(\bar{\sigma}, \bar{\sigma}')) $.  (3)  For any ordered strand pair
$ (\sigma, \sigma') $, we have $ y_{(\bar{\sigma}, \bar{\sigma}')} =
y_{(\sigma, \sigma')} $.

Because taking the complement of a strand essentially amounts to a relabelling 
of the bases and a change in the direction in which the strand is read, there 
is no reason to assume that conditions (1) - (3) should not hold in general.  
Indeed, cases where properties (1) - (3) do not hold indicate a 
strand asymmetry, a condition which results from specific, and 
presumably non-generic, base orderings.

If we assume that the fitness and ``mutation'' landscapes are chosen so
that properties (1) and (2) are met, then if our population initially
satisfies property (3) (obtained with a lesion-free population, for example),
it is possible to show that property (3) holds for all time.  The
proof of this is similar to the proof of the analogous statement for
quasispecies dynamics with imperfect lesion repair \cite{LESFULL}, and will
therefore be omitted here.

When lesion repair is perfect, then an initially lesion-free population
remains lesion free.  In this case we have,
\begin{eqnarray}
\frac{d y_{(\sigma, \bar{\sigma})}}{dt} 
& = &
-\kappa_{(\sigma, \bar{\sigma})} y_{(\sigma, \bar{\sigma})} 
\nonumber \\
&   &
+ \frac{1}{2} \sum_{(\sigma', \bar{\sigma}')}
\kappa_{(\sigma', \bar{\sigma}')} y_{(\sigma', \bar{\sigma}')}
p((\sigma', \bar{\sigma}'), (\sigma, \bar{\sigma})) 
\nonumber \\
&   &
+ \frac{1}{2} \sum_{(\sigma', \bar{\sigma}')}
\kappa_{(\bar{\sigma}', \sigma')} y_{(\bar{\sigma}', \bar{\sigma}')}
p((\bar{\sigma}', \sigma'), (\bar{\sigma}, \sigma))
\nonumber \\
& = &
-\kappa_{(\sigma, \bar{\sigma})} y_{(\sigma, \bar{\sigma})}
\nonumber \\
&   &
+ \frac{1}{2} \sum_{(\sigma', \bar{\sigma}')}
\kappa_{(\sigma', \bar{\sigma}')} y_{(\sigma', \bar{\sigma}')}
p((\sigma', \bar{\sigma}'), (\sigma, \bar{\sigma}))
\nonumber \\
&   &
+ \frac{1}{2} \sum_{(\sigma', \bar{\sigma}')}
\kappa_{(\sigma', \bar{\sigma}')} y_{(\sigma', \bar{\sigma}')}
p((\sigma', \bar{\sigma}'), (\sigma, \bar{\sigma}))
\nonumber \\
& = &
-\kappa_{(\sigma, \bar{\sigma})} y_{(\sigma, \bar{\sigma})}
\nonumber \\
&   &
+ \sum_{(\sigma', \bar{\sigma}')}
\kappa_{(\sigma', \bar{\sigma}')} y_{(\sigma', \bar{\sigma}')}
p((\sigma', \bar{\sigma}'), (\sigma, \bar{\sigma}))
\nonumber \\
\end{eqnarray}
which coincides with the immortal strand equations.

\section{The ``Master-Genome'' Fitness Landscape}

\subsection{Infinite sequence length equations}

Following the derivation of the quasispecies equations with imperfect lesion
repair \cite{LESFULL}, we will now develop the infinite sequence length
equations for a class of fitness landscapes defined by a ``master'' genome
$ \{\sigma_0, \bar{\sigma}_0\} $.  For simplicity, we assume that
$ \epsilon_{\{\sigma, \sigma'\}} $ and $ \lambda_{\{\sigma, \sigma'\}} $
are genome independent, and may respectively be denoted by $ \epsilon $
and $ \lambda $. 

Following the convention used with quasispecies dynamics, we derive the 
infinite sequence length equations with $ \mu \equiv L\epsilon $ held 
constant.  This is equivalent to fixing the the genome replication fidelity, 
given by $ e^{-\mu} $, in the limit of infinite sequence length.

The derivation of the infinite sequence length equations from the finite
sequence length equations for stem cell division parallels the derivation
of the infinite sequence length equations for semiconservative replication
with imperfect lesion repair.  We therefore refer the reader to \cite{LESFULL}
for details.  In this paper, we only provide the necessary definitions for
understanding the final form of the infinite sequence length equations.

To begin, we note that the ``master'' genome $ (\sigma_0, \bar{\sigma}_0) $
gives rise to the ordered sequence pairs $ (\sigma_0, \bar{\sigma}_0) $ 
and $ (\bar{\sigma}_0, \sigma_0) $.  In the limit of infinite sequence
length, the two master strands $ \sigma_0 $ and $ \bar{\sigma}_0 $ become
infinitely separated from each other in Hamming distance, hence we may
regard $ (\sigma_0, \bar{\sigma}_0) $ and $ (\bar{\sigma}_0, \sigma_0) $
as infinitely separated from each other in the ordered sequence pair space.

We may therefore group all sequence pairs $ (\sigma, \sigma') $ into one
of three classes:  A sequence pair $ (\sigma, \sigma') $ is said to be of
the {\it first class} if $ D_H(\sigma, \sigma_0) $ and 
$ D_H(\sigma', \bar{\sigma}_0) $ are both finite.  A sequence pair $ (\sigma,
\sigma') $ is said to be of the {\it second class} if 
$ D_H(\sigma, \bar{\sigma}_0) $ and $ D_H(\sigma', \sigma_0) $ are both
finite.  Finally, a sequence pair not belonging to either one of the
first two classes is said to belong to the {\it third class}.

A given sequence pair $ (\sigma, \sigma') $ of the first class can be 
characterized by the four parameters, denoted $ l_C $, $ l_L $, $ l_R $,
and $ l_B $.  The first parameter, $ l_C $, denotes the number of 
positions where $ \sigma $ and $ \sigma' $ are complementary, yet differ
from the corresponding positions in $ \sigma_0 $ and $ \bar{\sigma}_0 $,
respectively.  The second parameter, $ l_L $, denotes the number of positions
where $ \sigma $ differs from $ \sigma_0 $, but the complementary positions
in $ \sigma' $ are equal to the corresponding ones in $ \bar{\sigma}_0 $.
The third parameter, $ l_R $, denotes the number of positions where
$ \sigma $ is equal to the ones in $ \sigma_0 $, but the complementary
positions in $ \sigma' $ differ from the corresponding ones in 
$ \bar{\sigma}_0 $.  Finally, the fourth parameter, $ l_B $, denotes the
number of positions where $ \sigma $ and $ \sigma' $ are not complementary,
and also differ from the corresponding positions in $ \sigma_0 $ and
$ \bar{\sigma}_0 $.  These definitions are illustrated in Figure 3 of
\cite{LESFULL}.  A sequence pair of the second class may be similarly
characterized (except $ \sigma_0 $ and $ \bar{\sigma}_0 $ are swapped in the
definitions given above).

We assume that the fitness of a given sequence pair of the first class is 
determined by $ l_C $, $ l_L $, $ l_R $, and $ l_B $, hence we may write
that $ \kappa_{(\sigma, \sigma')} = \kappa_{(l_C, l_L, l_R, l_B)} $.  The
fitness of a sequence pair $ (\sigma, \sigma') $ of the second class is
determined by noting that $ (\sigma', \sigma) $ is of the first class, and
that $ \kappa_{(\sigma, \sigma')} = \kappa_{(\sigma', \sigma)} $.  We take
the third class sequence pairs to be unviable, with a first-order growth
rate of $ 1 $.

We also assume that $ \kappa_{(l_C, l_L, l_R, l_B)} = \kappa_{(l_C, l_R,
l_L, l_B)} $.  This is a natural assumption to make if one assumes symmetry
between the two master strands.  In \cite{LESFULL}, we show that this
assumption implies that $ \kappa_{(\bar{\sigma}, \bar{\sigma}')} = 
\kappa_{(\sigma, \sigma')} $.

We allow our system to come to equilibrium starting from the initial
condition $ y_{(\sigma_0, \bar{\sigma}_0)} = y_{(\bar{\sigma}_0, \sigma_0)} = 
1/2 $.  This initial condition corresponds to an initially mutation-free stem 
cell population.

We may sum over the population fractions of all first class sequence pairs 
characterized by a given set of $ l_C $, $ l_L $, $ l_R $, and $ l_B $,
and reexpress the quasispecies dynamics in terms of these quantities.
We define $ z_{(l_C, l_L, l_R, l_B)} $ to be the total population fraction
of first class sequence pairs characterized by $ l_C $, $ l_L $, $ l_R $,
and $ l_B $.  We similarly define $ \bar{z}_{(l_C, l_L, l_R, l_B)} $
to be the total population fraction of second class sequence pairs
characterized by $ l_C $, $ l_L $, $ l_R $, $ l_B $.  Following the
derivation in \cite{LESFULL}, we then obtain,
\begin{widetext}
\begin{eqnarray}
\frac{d z_{(l_C, l_L, l_R, 0)}}{dt} 
& = &
-\kappa_{(l_C, l_L, l_R, 0)} z_{(l_C, l_L, l_R, 0)} 
\nonumber \\
&   &
+ \frac{1}{2} (\frac{1}{l_L!} (\mu (1 - \lambda))^{l_L} \delta_{l_R 0} +
               \frac{1}{l_R!} (\mu (1 - \lambda))^{l_R} \delta_{l_L 0})
e^{-\mu (1 - \lambda/2)} \times \nonumber \\
&   &
\sum_{l_C' = 0}^{l_C} \frac{1}{l_C'!} 
(\frac{\lambda \mu}{2})^{l_C'}
\sum_{l_1'' = 0}^{l_C - l_C'}\sum_{l_2'' = 0}^{\infty}
\kappa_{(l_1'', l_C - l_C' - l_1'', l_2'', 0)} 
z_{(l_1'', l_C - l_C' - l_1'', l_2'', 0)}
\end{eqnarray}
for random segregation, and 
\begin{eqnarray}
\frac{d z_{(l_C, 0, l_R, 0)}}{dt}
& = &
-\kappa_{(l_C, 0, l_R, 0)} z_{(l_C, 0, l_R, 0)} \nonumber \\
&   &
+ \frac{1}{l_R!} (\mu (1 - \lambda))^{l_R} e^{-\mu (1 - \lambda/2)}
\sum_{l_C' = 0}^{l_C} \frac{1}{l_C'!} (\frac{\lambda \mu}{2})^{l_C'}
\sum_{l_1'' = 0}^{\infty} \kappa_{(l_C - l_C', 0, l_1'', 0)} 
z_{(l_C - l_C', 0, l_2'', 0)}
\end{eqnarray}
for immortal strand co-segregation.  An analogous set of equations may
be derived for the $ \bar{z}_{(l_C, l_L, l_R, l_B)} $.  Using the
fact that $ y_{(\bar{\sigma}, \bar{\sigma}')} = y_{(\sigma, \sigma')} $
we have $ \bar{z}_{(l_C, l_L, l_R, l_B)} = z_{(l_C, l_L, l_R, l_B)} $.  
\end{widetext}

An interesting feature to note from comparison of these two equations
is that for random chromosome segregation, it is possible for $ l_L > 0 $,
while for immortal strand co-segregation, we have $ l_L = 0 $.  In the case
of random segregation, the ordered strand pairs $ (\sigma, \sigma') $
and $ (\sigma', \sigma) $ are equivalent, hence we have
$ \bar{z}_{(l_C, l_R, l_L, l_B)} = z_{(l_C, l_L, l_R, l_B)} $,
which implies that $ z_{(l_C, l_R, l_L, l_B)} = z_{(l_C, l_L, l_R, l_B)} $.
In the case of immortal strand co-segregation, the first strand of the ordered
strand pair represents the parent strand.  Because the parent strand differs
from $ \sigma_0 $ (or $ \bar{\sigma}_0 $ when looking at the $ \bar{z} $
equations) in only a finite number of positions, in the limit of
infinite sequence length the probability that a mismatch occurs where
the parent strand differs from $ \sigma_0 $ is $ 0 $.  Therefore, any
lesions that occur will be due to an error made in the daughter strand,
where the corresponding bases of the parent strand are identical to those
of $ \sigma_0 $.  Thus, $ l_L $ remains $ 0 $, but $ l_R $ can become 
positive.

Finally, from these equations it is possible to show that a population of 
adult stem cells will eventually degrade unless lesion repair is turned off 
and chromosome segregation occurs via the immortal strand mechanism.  For 
random chromosome segregation, a given stem cell will periodically retain an 
erroneous daughter strand, resulting in a steady degradation of the genome.
For immortal strand co-segregation with nonzero lesion repair efficiency,
mistakes in the daughter strands will periodically be communicated to the
parent strand via lesion repair.  The result is again a steady
degradation of the genome.

\subsection{Decay of the master-genome population}

We may derive a set of differential equations describing the 
decay of the master genome population.  We consider a fitness landscape
where the viable genomes have a first-order growth rate constant $ k_{+} $,
and the unviable genomes have a first-order growth rate constant
$ k_{-} < k_{+} $.  An ordered strand pair is taken to be viable if 
$ l_C \leq l_{C, max} $, and if $ l_L + l_R + l_B \leq l $.  Thus, an
ordered strand pair is viable if it has no more than $ l_{C, max} $
fixed mutations, and no more than $ l $ lesions.  Otherwise,
the strand pair is unviable.

Defining $ z_0 = z_{(0, 0, 0, 0)} $, $ z_1 = \sum_{l' = 0}^{l}
{z_{(0, 0, l', 0)}} $, and $ z_2 = \sum_{l' = 0}^{\infty}{z_{(0, 0, l', 0)}} $,
we obtain, for random segregation, that, 
\begin{eqnarray}
&   &
\frac{d z_0}{dt} = 
-k_{+} z_0 + \frac{1}{2} e^{-\mu (1 - \lambda/2)}
[(k_{+} - k_{-}) z_1 + k_{-} z_2] \nonumber \\
&   &
\frac{d z_1}{dt} = -k_{+} z_1 + \frac{1}{2} (1 + f_l(\mu, \lambda))
e^{-\mu (1 - \lambda/2)} \times \nonumber \\
&   &
[(k_{+} - k_{-}) z_1 + k_{-} z_2] \nonumber \\
&   &
\frac{d z_2}{dt} = -(1 - \frac{1}{2}(e^{-\mu (1 - \lambda/2)} +
e^{-\mu \lambda/2})) \times \nonumber \\
&   &
[(k_{+} - k_{-}) z_1 + k_{-} z_2]
\end{eqnarray}

For immortal strand co-segregation, we obtain,
\begin{eqnarray}
&   &
\frac{d z_0}{dt} = -k_{+} z_0 + e^{-\mu (1 - \lambda/2)}
[(k_{+} - k_{-}) z_1 + k_{-} z_2]
\nonumber \\
&   &
\frac{d z_1}{dt} = -k_{+} z_1 + f_l(\mu, \lambda) e^{-\mu (1 - \lambda/2)}
[(k_{+} - k_{-}) z_1 + k_{-} z_2] \nonumber \\
&   &
\frac{d z_2}{dt} = -(1 - e^{-\mu \lambda/2})[(k_{+} - k_{-}) z_1 + k_{-} z_2] 
\end{eqnarray}

We may solve Eqs. (13) and (14) using standard numerical methods, for 
the initial condition $ z_0 = z_1 = z_2 = 1/2 $.  This corresponds 
to an initial stem cell population consisting entirely of the master
genome genotype.

In Figure 3 we show a comparison of the numerical solution of Eqs. (13)
and (14) with the results of stochastic simulations of dividing stem
cells.  The lesion repair probability $ \lambda $ is taken to be $ 0.5 $
in this case.

\begin{figure}
\includegraphics[width = 0.7\linewidth, angle = -90]{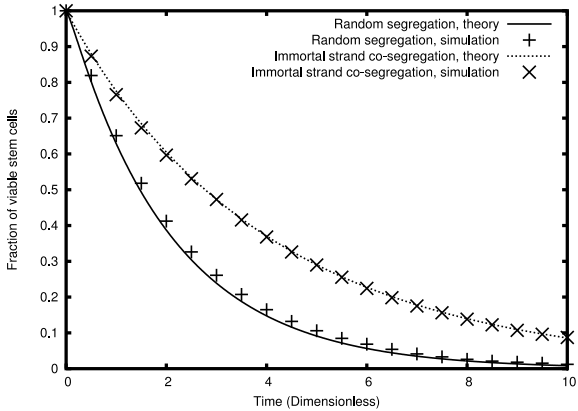}
\caption{Comparison of theory and simulation for a population of $ 10,000 $
stem cells with genomes of sequence length $ 20 $.  We assume $ k_{+} = 10 $, 
$ k_{-} = 1 $, $ \mu = 0.1 $, and $ l = 1 $.  We iterated in time steps of 
length $ 0.001 $ out to a time of $ 10 $.}
\end{figure}

\section{Optimal Lesion Repair Probabilities}

We can use Eq. (14) to determine the optimal lesion repair probability for
preserving the stem cell line out to a given time $ T $.  We use $ z_0 $
as our measure for the extent of the preservation of the stem cell line.
The higher the value of $ z_0 $, the better the stem cell line is preserved.
To this end, for simplicity, we also take $ k_{-} = 0 $, i.e., we assume
that unviable stem cells do not replicate at all.  We also rescale the
time by defining $ \tau = k_{+} t $.  We then obtain,
\begin{equation}
z_0(\tau) = \frac{1}{2} e^{-\tau}[1 + \frac{1}{f_l(\mu, \lambda)}
(e^{f_l(\mu, \lambda) exp(-\mu (1 - \lambda/2)) \tau} - 1)]
\end{equation}
Therefore, maximizing $ z_0(T) $ is equivalent to maximizing 
$ g_l(\lambda; \mu, T) \equiv (e^{f_l(\mu, \lambda) 
\exp(-\mu (1 - \lambda/2)) T} - 1)/f_l(\mu, \lambda) $.

It is instructive to consider the behavior of $ g_l $ for $ l = 0 $ and
$ l = \infty $.  For $ l = 0 $, we have $ g_0 = 
e^{\exp(-\mu (1 - \lambda/2)) T} - 1 $, which is clearly maximized for
any $ \mu $ and $ T $ when $ \lambda = 1 $.  This makes sense because, when
$ l = 0 $, then any lesion renders the stem cell unviable.  Preserving
the information in the parent strand by reducing the lesion repair efficiency
does not help maintain the population of master genomes, since an unviable
stem cell does not replicate.  Therefore, in this case, it is optimal to
make lesion repair maximally efficient, thereby reducing the overall mutation
rate away from the master genome.

For imperfect lesion repair to allow for better preservation of the stem cell
population within our model, we must therefore assume that $ l > 0 $.  While 
typical values of $ l $ for cellular organisms are not available (the matter
is also complicated by additional repair mechanisms such as SOS response), 
we may note that the smaller the value of $ \mu $, the fewer errors are
made during replication (an average of $ \mu $ are made).  Thus, in practice, 
for small $ \mu $, one may assume that $ l = \infty $, since a large number of 
lesions will not be produced in any case (mathematically, this is equivalent 
to the observation that the series $ \{f_l(\mu, \lambda)\} $ converges to 
$ f_{\infty}(\mu, \lambda) = e^{\mu (1 - \lambda)} $ more quickly at 
smaller values of $ \mu $ than at larger values of $ \mu $).  Since cells
have various error correction mechanisms which keep the overall number
of replication errors to on the order of $ 1 $ or less per replication
cycle, the assumption that $ l = \infty $ seems to be a reasonable one,
and will be used here.  

For $ l = \infty $, we then have $ g_{\infty} =
e^{-\mu (1 - \lambda)} (e^{\exp(-\mu \lambda/2) T} - 1) $.  For a given
$ \mu $ and $ T $, we define $ y = e^{-\mu \lambda/2} T $, giving
$ g_{\infty} = e^{-\mu} T^2 (e^y - 1)/y^2 $.  The function $ (e^y - 1)/y^2 $
goes to $ \infty $ at $ y = 0 $ and $ y = \infty $.  It has a unique point
where its derivative vanishes, corresponding to a global minimum.  Thus, on 
any given interval, the maximum value of $ (e^y - 1)/y^2 $ occurs at one of 
the endpoints.  In particular, this implies that $ g_{\infty} $ is maximized 
for a given $ \mu $ and $ T $ at either $ \lambda = 0 $ or $ \lambda = 1 $.

To determine whether the optimal $ \lambda $ is $ 0 $ or $ 1 $ for given
values of $ \mu $ and $ T $, we note that $ \lambda = 0 $ corresponds
to $ y = T $, while $ \lambda = 1 $ corresponds to $ y = e^{-\mu/2} T $.
The minimum value of $ (e^y - 1)/y^2 $ occurs before $ y = 2 $, hence,
once $ e^{-\mu/2} T > 2 $, $ (e^y - 1)/y^2 $ becomes monotone increasing
on $ [e^{-\mu/2} T, T] $, so that $ g_{\infty} $ is maximized for 
$ \lambda = 0 $.  For human cells, the genome length is of the order of 
$ 3 \times 10^9 $ base pairs, giving $ \mu \approx 3 $ \cite{VOET}.  
Therefore, if $ T > 2 e^{-3/2} \approx 9 $, then optimal preservation of the 
stem cell line occurs for $ \lambda = 0 $.  Current estimates place the 
number of adult stem cell divisions in the human colon over a human lifetime 
at around  $ 5,000 $ \cite{POTTEN2}.  In our rescaled time coordinates, this 
gives $ T = 5,000 >> 9 $.  Clearly then, to optimally preserve the stem cell 
line, our model indicates that lesion repair should be turned off during cell 
division.

We should note that, at short times, it is optimal to keep $ \lambda = 1 $,
indepedent of $ l $ (this can be shown by expanding $ g_l $ out to first-order
in $ T $, and optimizing).  Also, for finite values of $ l $, it is possible
to show that, at sufficiently long times, the optimal lesion repair 
efficiency can be made arbitrarily close to $ 1 $ by making the mutation
rate $ \mu $ arbitrarily large.  This makes sense, because, at high mutation
rates, it is necessary to prevent the formation of more than $ l $ lesions
during replication, which renders the adult stem cell unviable.

For our purposes, however, the $ l = \infty $ simplification seems
appropriate, since it is reasonable to assume that $ \mu = 3 $ is 
considerably less than the number of mismatches which a human adult stem
cell can tolerate before becoming unviable.

It is important to note that, by lesion repair, we specifically refer
to mismatched base-pairs along the DNA chain.  The underlying assumption,
however, is that each of the bases are chosen from one of the four standard
bases (A, T, G, C).  Thus, when considering lesions in this model, we are
not considering lesions caused by chemical modifications of bases, due to,
for example, radiation or oxidative damage.  In principle, these lesions can 
be correctly repaired, assuming that the damage is localized to only one of
the strands, because the chemical changes to the bases allows the cellular
repair mechanisms to determine on which strand the lesion is present.

Thus, in determining that for human stem cells, lesion repair should
be turned off during cell division, we mean that mismatches along the
DNA genome should be left alone, so as not to risk fixing a mutation
in both strands.  

While it is possible that distinct cellular mechanisms exist for repairing
postreplication mismatches and lesions due to DNA damage, it is also
possible that both types of modifications to a DNA genome are handled
by the same repair pathways (Nucleotide Excision Repair, for instance
\cite{VOET}).  Thus, it is possible that the way by which adult stem cells
suppress correction of mismatches along the DNA chain is by a general
suppression of lesion-repair.  In this case, adult stem cells should be 
more susceptible to the effects of agents which can damage DNA.  This 
increased susceptibility to DNA damage has been hypothesized by Cairns
\cite{CAIRNS2}, and does indeed appear to be a property of adult stem
cells \cite{CAIRNS2}.

\section{Conclusions}

This paper developed a set of ordinary differential equations describing
the evolutionary dynamics of a population of adult stem cells.  For 
simplicity, we considered stem cell genomes consisting of a single 
double-stranded DNA molecule, i.e., one chromosome.  

We considered two possible mechanisms of chromosome segregation.  In the first 
case, we assumed that chromosomes randomly segregate into the adult stem cell 
and undifferentiated tissue cell.  In the second case, we assumed that the 
stem cell retains the chromosome containing the oldest DNA strand of the 
genome.  This co-segregation mechanism, termed the immortal strand hypothesis, 
was originally proposed by Cairns in 1975 \cite{CAIRNS1} as a mechanism
by which stem cells preserve the integrity of their genomes.

For the case of random segregation, we derived a set of equations analogous
to the quasispecies equations for semiconservative replication with imperfect
lesion repair.  In particular, the ordered strand pair formalism developed
in \cite{LESFULL} was used.  

For immortal strand co-segregation, we showed that an analogous ordered strand
pair formalism is possible, though in contrast to random segregation, the 
labelling of parent and daughter strands leads to a canonical method for 
constructing an ordered strand pair from a given genome.  This results in
a different set of equations describing the dynamics over the space of
ordered strand pairs.

Following the approach taken with the semiconservative quasispecies equations
with imperfect lesion repair \cite{LESFULL}, we developed the infinite
sequence length equations for the stem cell population, assuming a 
fitness landscape defined by a master-genome.  From both the random and
immortal strand equations it is readily shown that immortal strand
segregation with imperfect lesion repair helps to maintain a population of 
stem cells.

From the infinite sequence length equations, we obtained the differential 
equations governing the decay of the master genome population, and developed
a criterion for determining the optimal lesion repair probability for 
maximizing the population of stem cells with the genotype defined by the
master genome.  Based on parameters for human stem cells, we predict that
lesion repair should be completely turned off in adult human stem cells.
This result, of course, is in the end a prediction made by a highly
simplified model, and needs to be experimentally tested.  Furthermore,
because it appears that postreplication mismatches and lesions due to
DNA damage are repaired by the same biochemical pathways \cite{CAIRNS2}, 
future research will need to explicitly incorporate DNA damage in order
to refine our estimate for optimal lesion repair efficiency in adult
stem cells.  Nevertheless, despite the simplifying assumptions made in
this work, we regard this paper as an important first step toward a
quantitative modeling of stem cell evolutionary dynamics.

In this paper, we assumed that the stem cell and tissue genomes consist of 
only one chromosome.  While one chromosome is sufficient for studying 
immortal strand co-segregation, in reality vertebrate cells contain numerous 
chromosomes.  Furthermore, it is known that certain free living organisms, 
such as {\it Saccharomyces cerevisiae} variants (Baker's yeast), segregate 
chromosomes according to the immortal strand mechanism \cite{TANAKA}.  For 
single-chromosome genomes, the immortal strand mechanism cannot be applied 
to free living cells, since there is no qualitative distinction between the 
two daughter cells (such as ``stem'' and ``tissue'').  However, with multiple 
chromosomes, it is possible for asymmetric segregation to occur so that one
of the daughter cells retains the chromosomes with the oldest DNA strands.
Thus, the study of immortal strand co-segregation for multiple chromosome
genomes is an important extension of the model presented here
and the imperfect lesion repair quasispecies equations presented in
\cite{LESFULL}.

\begin{acknowledgments}

This research was supported by the National Institutes of Health.  

\end{acknowledgments}

\end{document}